\documentclass[journal,twoside,final]{IEEEtran}
\usepackage{graphicx}
\usepackage{amsmath}
\usepackage{latexsym}
\usepackage{graphicx}
\usepackage{bm}
\usepackage{amssymb}
\usepackage{array}
\usepackage{setspace}
\usepackage{fancyhdr}
\usepackage{amsmath, amssymb, epsfig}
\usepackage{float}
\usepackage{subcaption}

\usepackage{color}

\usepackage{csquotes}
\MakeOuterQuote{"}

\makeatletter
\newcommand{\vast}{\bBigg@{3.2}}
\newcommand{\Vast}{\bBigg@{4.2}}
\makeatother

\begin{document}
%
\title{\linespread{1} Mixing RIS-Assisted Sources with FSO Link}

\author{Anas~M.~Salhab\authorrefmark{1},~\IEEEmembership{Senior~Member,~IEEE} and Liang Yang\authorrefmark{2}\\
\thanks{\authorrefmark{1} M. Salhab is with the Department
of Electrical Engineering, King Fahd University of Petroleum \&
Minerals, Dhahran 31261, Saudi Arabia (e-mail: salhab@kfupm.edu.sa).}

\thanks{\authorrefmark{2} Yang is with the College of Computer Science and Electronic
Engineering, Hunan University, Changsha 410082, China, and also with the
State Key Laboratory of Integrated Services Networks, Xidian University,
Xi’an 710126, China (e-mail: liangy@hnu.edu.cn).}}



\maketitle{}

\begin{abstract}
This letter proposes and evaluates the performance of reconfigurable 
intelligent surface (RIS)-assisted source 
mixed radio frequency (RF)/free space optical
(FSO) relay network with opportunistic source scheduling. Closed-form analytical approximations 
are derived for the outage probability and average symbol error probability (ASEP)
assuming Rayleigh and Gamma-Gamma fading models for the RF and FSO
channels, respectively. Moreover, the system is studied at the high signal-to-noise
ratio (SNR) regime, whereby the diversity order and coding gain are
derived and analyzed. The results illustrate that the system performance is
dominated by the worst hop and that the
diversity order is equal to $G_{d}=\mathrm{min}(KN,\alpha,\beta,\zeta^{2})$, where $K$ is the number of RF sources, $N$ is the number of reflecting elements at each source, $\alpha$ and $\beta$ are the atmospheric turbulence parameters of the FSO link, and $\zeta^{2}$ is a measure for the alignment quality of the FSO source and receiver. In addition, findings show that for the same $G_{d}$, $N$ is more impactful on the system performance than $K$ through the coding gain $G_{c}$.
\end{abstract}



%
\IEEEpeerreviewmaketitle

\section{Introduction}
Free space optical (FSO) communication represents an efficient means to deal with the "last-mile"
problem in wireless networks \cite{Kedar}. In such systems, the
data transmission takes place between an optical transmitter and a
receiver located, for example, on high buildings, separated by
several hundred meters. Relay networks are also efficient in handling the
multipath fading problem in wireless networks \cite{Laneman}.
A mixture of relay and FSO networks has been widely introduced
to increase the coverage distance of FSO
networks, which is usually limited to a few hundred meters due to atmospheric turbulence conditions
\cite{Lee}. The mixed radio frequency (RF)/FSO relaying can be
used to multiplex multiple RF users into a single FSO link
\cite{Ansari1}.

Reconfigurable intelligent surfaces (RISs) have recently attracted a noticeable attention as a promising technique for future wireless communication networks. An RIS is an artificial surface, made of electromagnetic
material, that is capable of customizing the propagation of
the radio waves impinging upon it \cite{Renzo1}. 
It represents a new low-cost and less complicated solution to realize wireless
communication with high energy and spectrum efficiencies \cite{Basar1}.

In \cite{Wu}, it has been shown that RIS has better performance than
conventional massive multiple-input multiple-output
systems as well as better performance than multi-antenna
amplify-and-forward relaying networks with smaller
number of antennas, while reducing the system complexity and
cost. Authors in \cite{Yang3} considered a dual-hop mixed FSO/RF network, where an RIS has been used on the second hop to forward the source message to destination. Most of the existing works considered the scenario, where an RIS is used to replace relays. There exists another important scenario, where the RIS could be used as part of the transmitter along with the RF signal generator to help the source in its data communications \cite{Renzo1}, \cite{Yang6}. Recently, the idea of using an RIS as a transmitter was validated with the aid of a testbed platform in \cite{Tang}.

In this letter, we consider this scenario where multiple RIS-assisted sources are connected to a relay node via RF links and the relay is connected to a destination via a FSO link. Opportunistic source scheduling is used to select among RF sources. Closed-form analytical approximations are derived for the outage probability and average symbol error probability (ASEP) assuming Rayleigh and Gamma-Gamma fading channels for the RF and FSO links, respectively. The system is also studied at the high signal-to-noise ratio (SNR) regime, where the diversity order and coding gain are provided.







\section{System and Channel Models}\label{SCMs}
Consider a dual-hop mixed RF/FSO relay network consisted of
$K$ RIS-assisted sources on the first hop ${\sf U}_{k}$
($k=1,\ldots,K$), one decode-and-forward (DF) relay ${\sf R}$, and
one destination ${\sf D}$. The sources are connected with the
relay through RF links, and the relay is connected with a destination
through a FSO link. It is assumed that each
source is equipped with $N$ reflecting elements, the relay is
equipped with a single antenna and a single photo-aperture
transmitter, and the destination is equipped
with a single antenna. The communication happens over two phases: selected source
${\sf U}_{\mathrm{Sel}}$ $\rightarrow$ ${\sf R}$ and ${\sf R} \rightarrow {\sf
D}$. The received signal at ${\sf R}$ from the
$k^{\mathrm{th}}$ source can be expressed as
\begin{align}\label{Eq.1}
y_{k,{\sf r}}=\sqrt{P_{k}}\sum_{i=1}^{N}h_{k,{\sf r},i}e^{j\phi_{k,{\sf r},i}}x_{k,{\sf
r}}+n_{{\sf r}},
\end{align}
where $P_{k}$ is the transmit power of the $k^{\mathrm{th}}$ source, $N$ is the number of reflecting elements,
$h_{k,{\sf r},i}=\alpha_{k,{\sf r},i}e^{-j\theta_{k,{\sf r},i}}$ is the channel coefficient of the $i^{\mathrm{th}}$ reflecting element at ${\sf
U}_{k}$ and ${\sf R}$, where $\alpha_{k,{\sf r},i}$ is the amplitude and $\theta_{k,{\sf r},i}$ is the phase shift, $\phi_{k,{\sf r},i}$ is the adjustable phase induced by the $i^{\mathrm{th}}$ reflecting
element, $x_{k}$ is the
transmitted symbol of ${\sf U}_{k}$ with $\mathbb{E}\{|{x_{k}}|^{2}\}=1$, and $n_{{\sf r}}\thicksim\mathcal{N}(0,
N_{0,{\sf r}})$ is an additive white Gaussian noise (AWGN)
term, where $\mathbb{E}\{\cdot\}$ is the mathematical expectation.
Using (\ref{Eq.1}), the SNR at ${\sf R}$ due to ${\sf U}_{k}$
can be written as \cite{Basar1}
\begin{align}\label{Eq.2}
\gamma_{{\sf U}_{k},{\sf R}}&=\frac{P_{k}}{N_{0,{\sf
r}}}\left(\sum_{i=1}^{N-1}|\alpha_{k,{\sf r},i}|\right)^{2}.
\end{align}

Using opportunistic source scheduling, the source selection takes place such that
$\gamma_{{\sf U}_{\mathrm{Sel}},{\sf
R}}=\underset{\substack{k}}{\
\mathrm{max}}\left\{\gamma_{{\sf U}_{k},{\sf R}}\right\}$. The
subcarrier intensity modulation (SIM) scheme is employed at ${\sf
R}$, where a standard RF coherent/noncoherent modulator and
demodulator can be used for transmitting and recovering the source
data \cite{Salhab_IEEE_JLT_2016}. At ${\sf R}$, after
filtering by a bandpass filter, a direct current (DC) bias is
added to the filtered RF signal to ensure that the optical signal
is non-negative. Then the biased signal is sent to a continuous
wave laser driver. The retransmitted optical signal at ${\sf
R}$ is written as \cite{Lee}
\begin{equation}\label{Eq.1n1}
y_{{\sf
r}}^{\mathrm{Opt}}=\sqrt{P_{\mathrm{Opt}}}(1+\mathcal{M}y_{\mathrm{Sel},{\sf
r}}),
\end{equation}
where $P_{\mathrm{Opt}}$ denotes the average transmitted optical
power and it is related to the relay electrical power $P_{\sf r}$
by the electrical-to-optical conversion efficiency $\eta_{1}$ as
$P_{\mathrm{Opt}}=\eta_{1}P_{{\sf r}}$, $\mathcal{M}$ denotes
the modulation index, and $y_{\mathrm{Sel},{\sf r}}$ is the RF
received signal at ${\sf R}$ from the selected source. The
optical signal at ${\sf D}$ received from ${\sf R}$ at the
second phase of communication can be expressed as
\begin{align}\label{Eq.3n1}
y_{{\sf r},{\sf D}}=&g_{{\sf r},{\sf
d}}\left\{\sqrt{P_{\mathrm{Opt}}}\left[1+\mathcal{M}\left(\sqrt{P_{\mathrm{Sel}}}\sum_{i=1}^{N}\alpha_{\mathrm{Sel},{\sf
r},i}x_{\mathrm{Sel},{\sf r}}+n_{{\sf
r}}\right)\right]\right\}\nonumber\\
&+n_{{\sf d}},
\end{align}
where $n_{{\sf r}}\thicksim\mathcal{N}(0, N_{0,{\sf d}})$
is an AWGN term at ${\sf D}$. Moreover, the channel
coefficient of the ${\sf R}$$\rightarrow{\sf D}$ link,
which is given by $g_{{\sf r},{\sf d}}$ is modelled as
$g_{{\sf r},{\sf d}}=g_{a}g_{f}$, where $g_{a}$ and
$g_{f}$ are the average gain and the fading gain of the FSO link,
respectively \cite{Zhang}. When the DC component is filtered out
at ${\sf D}$ and an optical-to-electrical conversion is
performed and assuming $\mathcal{M}=1$, the received signal can be
expressed as
\begin{align}\label{Eq.3qa}
y_{{\sf r},{\sf d}}=g_{{\sf r},{\sf
d}}\sqrt{P_{\mathrm{Ele}}}\left(\sqrt{P_{\mathrm{Sel}}}\sum_{i=1}^{N}\alpha_{\mathrm{Sel},{\sf
r},i}x_{\mathrm{Sel},{\sf r}}+n_{{\sf r}}\right)+n_{{\sf
d}},
\end{align}
where $P_{\mathrm{Ele}}=\eta_{2} P_{\mathrm{Opt}}=\eta_{1}\eta_{2}
P_{{\sf r}}$ is the electrical power received at ${\sf D}$
with $\eta_{2}$ is the optical-to-electrical conversion efficiency
and $P_{{\sf r}}$ is the transmit power at ${\sf R}$. From
\eqref{Eq.3qa}, the end-to-end (e2e) SNR at ${\sf D}$ can be written using the
standard approximation $\gamma_{{\sf D}}\cong
\mathrm{min}(\gamma_{{\sf U}_{\mathrm{Sel}},{\sf
R}},\gamma_{{\sf R},{\sf D}})$ as \cite{Ansari1}
\begin{align}\label{Eq.6qa}
\gamma_{{\sf D}}=\frac{\gamma_{{\sf U}_{\mathrm{Sel}},{\sf
R}}\gamma_{{\sf R},{\sf D}}}{\gamma_{{\sf
U}_{\mathrm{Sel}},{\sf R}}+\gamma_{{\sf R},{\sf D}}},
\end{align}
where $\gamma_{{\sf U}_{\mathrm{Sel}},{\sf R}}=\frac{P_{
\mathrm{Sel}}}{N_{0,{\sf r}}}\left(\sum_{i=1}^{N}|\alpha_{\mathrm{Sel},{\sf
r},i}|\right)^{2}$ and $\gamma_{{\sf R},{\sf
D}}=\frac{\eta_{1}\eta_{2}P_{{\sf r}}}{N_{0,{\sf
d}}}|g_{{\sf r},{\sf d}}|^{2}$, which can be also
represented by $\gamma_{{\sf R},{\sf
D}}=\frac{P_{\mathrm{Ele}}}{N_{0,{\sf d}}}|g_{{\sf
r},{\sf d}}|^{2}=\frac{\eta_{2}
P_{\mathrm{Opt}}}{N_{0,{\sf d}}}|g_{{\sf r},{\sf
d}}|^{2}$. 


We assume that the channel coefficients between the sources and the relay ${\sf R}$ $\{{\alpha_{k,{\sf r},i}}, k=1,...,K; i=1,...,N\}$ to be Rayleigh distributed with mean $\frac{\sqrt{\pi}}{2}$ and variance $\frac{(4-\pi)}{4}$. That is, their mean powers $\mathbb{E}\{|\alpha_{k,{\sf r},i}|^{2}\}=1$. The cumulative distribution function (CDF) of $\gamma_{{\sf U}_{\mathrm{Sel}},{\sf R}}$ assuming independent identically distributed (i.i.d.) channels $\{\bar{\gamma}_{1,{\sf r}}=\bar{\gamma}_{2,{\sf r}}=...=\bar{\gamma}_{K,{\sf r}}=\bar{\gamma}_{u,{\sf r}}=\frac{P_{u}}{N_{0,{\sf r}}}\}$ is given by \cite{Yang6}
\begin{align}\label{sese}
F_{\gamma_{{\sf U}_{u},{\sf R}}}(\gamma) = 1- e^{-\frac{\gamma}{C\bar{\gamma}_{u,{\sf r}}}} \sum_{i = 0}^{N - 1} \frac{\gamma^i}{(C\bar{\gamma}_{u,{\sf r}})^{i} i!},
\end{align}
where $C = 1 + (N - 1)\Gamma^{2}\left(\frac{3}{2} \right)$, with $\Gamma(.)$ is the Gamma
function as defined in \cite[Eq. (8.310)]{Grad.}.

We assume that the FSO link experiences a unified Gamma-Gamma fading model
with pointing errors effect whose SNR probability density function (PDF) is given by
\cite{Ansari1}, \cite{Ansari3}
\begin{align}\label{Eq.5}
f_{\gamma_{{\sf R},{\sf
D}}}(\gamma)=\frac{\zeta^{2}}{r\gamma\Gamma(\alpha)\Gamma(\beta)}
\mathrm{G}^{3,0}_{1,3}\left[\alpha\beta\left(\frac{\gamma}{\bar{\gamma}_{{\sf r},{\sf d}}}\right)^{\frac{1}{r}} \bigg|\substack{\zeta^{2}+1\\
\\\zeta^{2},\alpha,\beta}\right],
\end{align}
where $\zeta$ is the ratio between the equivalent beam radius at
the receiver and the pointing error displacement standard
deviation (jitter) at the receiver (i.e. when
$\zeta\rightarrow\infty$, we get the non-pointing error case)
\cite{Ansari1}, $r$ is the parameter defining the type of
detection technique (i.e. $r=1$ represents heterodyne detection
and $r=2$ represents intensity modulation (IM)/direct detection
(DD)), $\alpha$ and $\beta$ are the fading parameters related to
the atmospheric turbulence conditions \cite{AnasP}, $\bar{\gamma}_{{\sf r},{\sf
d}}=\frac{P_{\mathrm{Ele}}}{N_{0,{\sf
r}}}\mathbb{E}\{|{g_{{\sf r},{\sf
d}}}|^{2}\}=\frac{\eta_{2} P_{\mathrm{Opt}}}{N_{0,{\sf
r}}}\mathbb{E}\{|{g_{{\sf r},{\sf
d}}}|^{2}\}=\frac{\eta_{1}\eta_{2}P_{{\sf r}}}{N_{0,{\sf
r}}}\mathbb{E}\{|{g_{{\sf r},{\sf
d}}}|^{2}\}=\frac{\eta_{1}\eta_{2}P_{{\sf r}}}{N_{0,{\sf
d}}}\mu_{{\sf r},{\sf d}}$, and $\mathrm{G}(.)$ is the
Meijer G-function as defined in \cite[Eq. (9.301)]{Grad.}.  

\section{Performance Analysis}\label{EPA}

\subsection{Outage Probability}
The outage probability is defined as $P_{\sf
out}=\mathrm{Pr}\left[\gamma_{\sf D}\leq\gamma_{\sf out}\right]$,
where $\mathrm{Pr}[.]$ is the probability operation and
$\gamma_{\sf out}$ is a fixed outage threshold. The
CDF of $\gamma_{\sf D}$ can be
written as \cite{Salhab_IEEE_JLT_2016}
\begin{align}\label{Eq.7}
F_{\gamma_{\sf D}}(\gamma)&=1-(1-F_{\gamma_{{\sf
U}_{\mathrm{Sel}},{\sf R}}}(\gamma))(1-F_{\gamma_{{\sf
R},{\sf D}}}(\gamma))\nonumber\\
&=F_{\gamma_{{\sf
U}_{\mathrm{Sel}},{\sf R}}}(\gamma)+F_{\gamma_{{\sf
R},{\sf D}}}(\gamma)-F_{\gamma_{{\sf
U}_{\mathrm{Sel}},{\sf R}}}(\gamma)F_{\gamma_{{\sf
R},{\sf D}}}(\gamma),
\end{align}
where $F_{\gamma_{{\sf U}_{\mathrm{Sel}},{\sf R}}}(\gamma)$,
$F_{\gamma_{{\sf R},{\sf D}}}(\gamma)$ are the
CDFs of first hop and second hop SNRs, respectively.

\subsubsection{First Hop Link} Using opportunistic scheduling, the CDF $F_{\gamma_{{\sf U}_{\mathrm{Sel}},{\sf
R}}}(\gamma)$ can be written for i.i.d. sources channels as
\begin{align}\label{Eq.8}
F_{\gamma_{{\sf U}_{\mathrm{Sel}},{\sf R}}}(\gamma)=&\left(F_{\gamma_{{\sf U}_{u},{\sf R}}}(\gamma)\right)^{K},
\end{align}
where $F_{\gamma_{{\sf U}_{u},{\sf R}}}(\gamma)$ is the CDF provided by \eqref{sese}. Upon substituting \eqref{sese} in \eqref{Eq.8} and using the Binomial rule, we get
\begin{align}\label{Eq.10}
F_{\gamma_{{\sf U}_{\mathrm{Sel}},{\sf
R}}}(\gamma)=&\sum_{k=0}^{K}\begin{pmatrix}
K \\ k \end{pmatrix} (-1)^{k}e^{-\frac{k}{C\bar{\gamma}_{u,{\sf r}}}\gamma}\sum_{j_{1}=0}^{N-1}...\sum_{j_{k}=0}^{N-1}\gamma^{\sum_{n=1}^{k}j_{n}}\nonumber\\
&\times\frac{1}{(C\bar{\gamma}_{u,{\sf r}})^{\sum_{n=1}^{k}j_{n}}\prod_{n=1}^{k}j_{n}!}.
\end{align}

\subsubsection{Second Hop Link}
The CDF $F_{\gamma_{{\sf R},{\sf D}}}(\gamma)$ can be
obtained by integrating the PDF in \eqref{Eq.5} using
$\int_{0}^{\gamma}f_{\gamma_{{\sf R},{\sf D}}}(t)dt$ to
get \cite{Salhab_IEEE_JLT_2016}
\begin{align}\label{Eq.11}
F_{\gamma_{{\sf R},{\sf D}}}(\gamma)=A
\mathrm{G}^{3r,1}_{r+1,3r+1}\left[\frac{B}{\bar{\gamma}_{{\sf r},{\sf d}}}\gamma \bigg|\substack{1,\chi_{1}\\
\\\chi_{2},0}\right],
\end{align}
where
$A=\frac{r^{\alpha+\beta-2}\zeta^{2}}{(2\pi)^{r-1}\Gamma(\alpha)\Gamma(\beta)}$,
$B=\frac{(\alpha\beta)^{r}}{r^{2r}}$,
$\chi_{1}=\frac{\zeta^{2}+1}{r},\ldots,\frac{\zeta^{2}+r}{r}$
comprises of $r$ terms, and
$\chi_{2}=\frac{\zeta^{2}}{r},\ldots,\frac{\zeta^{2}+r-1}{r},
\frac{\alpha}{r},
\ldots,\frac{\alpha+r-1}{r},\frac{\beta}{r},\ldots,\frac{\beta+r-1}{r}$
comprises of $3r$ terms. Upon substituting \eqref{Eq.10} and \eqref{Eq.11} in \eqref{Eq.7} and after some simplifications, we
get \eqref{Eq.12}. By replacing
$\gamma$ by $\gamma_{\sf out}$, the outage probability is
obtained.

\begin{figure*}
\small
\begin{align}\label{Eq.12}
F_{\gamma_{\sf D}}(\gamma)=&\sum_{k=0}^{K}\begin{pmatrix}
K \\ k \end{pmatrix} (-1)^{k}e^{-\frac{k}{C\bar{\gamma}_{u,{\sf r}}}\gamma}\sum_{j_{1}=0}^{N-1}...\sum_{j_{k}=0}^{N-1}\gamma^{\sum_{n=1}^{k}j_{n}}\frac{(C\bar{\gamma}_{u,{\sf r}})^{-\sum_{n=1}^{k}j_{n}}}{\prod_{n=1}^{k}j_{n}!}\left(1-A
\mathrm{G}^{3r,1}_{r+1,3r+1}\left[\frac{B}{\bar{\gamma}_{{\sf r},{\sf d}}}\gamma \bigg|\substack{1,\chi_{1}\\
\\\chi_{2},0}\right]\right)+A
\mathrm{G}^{3r,1}_{r+1,3r+1}\left[\frac{B}{\bar{\gamma}_{{\sf r},{\sf d}}}\gamma \bigg|\substack{1,\chi_{1}\\
\\\chi_{2},0}\right].
\end{align}
\normalsize
 \hrule height 0.8pt
\end{figure*}

\begin{figure*}
\small
\begin{align}\label{Eq.14}
&\mathrm{ASEP}=\frac{a\sqrt{b}}{2\sqrt{\pi}}\Bigg\{\sum_{k=0}^{K}\begin{pmatrix}
K \\ k \end{pmatrix} (-1)^{k}\sum_{j_{1}=0}^{N-1}...\sum_{j_{k}=0}^{N-1}\frac{(C\bar{\gamma}_{u,{\sf r}})^{-\sum_{n=1}^{k}j_{n}}}{\prod_{n=1}^{k}j_{n}!\left(\frac{k}{C\bar{\gamma}_{u,{\sf r}}}+b\right)^{\delta}}\Bigg\{\Gamma\left(\delta\right)-A
\mathrm{G}^{3r,2}_{r+2,3r+1}\left[\frac{B}{\left(\frac{k}{C\bar{\gamma}_{u,{\sf r}}}+b\right)\bar{\gamma}_{{\sf r},{\sf d}}}\bigg|\substack{-\sum_{n=1}^{k}j_{n}+\frac{1}{2},1,\chi_{1}\\
\\\chi_{2},0}\right]\Bigg\}\nonumber\\
&+Ab^{-1/2}
\mathrm{G}^{3r,2}_{r+2,3r+1}\left[\frac{B}{b\bar{\gamma}_{{\sf r},{\sf d}}}\bigg|\substack{\frac{1}{2},1,\chi_{1}\\
\\\chi_{2},0}\right],\ \mathrm{where}\ \delta=\sum_{n=1}^{k}j_{n}+1/2.
\end{align}
\normalsize
 \hrule height 0.8pt
\end{figure*}

\subsection{Average Symbol Error Probability}
The ASEP is expressed in
terms of the CDF of $\gamma_{\sf D}$ as \cite{McKay}
\begin{align}\label{Eq.13}
\mathrm{ASEP}=\frac{a\sqrt{b}}{2\sqrt{\pi}}\int_{0}^{\infty}\frac{\exp\left(-b\gamma\right)}{\sqrt{\gamma}}F_{\gamma_{\sf
D}}(\gamma) d\gamma,
\end{align}
where $a$ and $b$ are modulation-specific parameters. Substituting
\eqref{Eq.12} in \eqref{Eq.13} and using \cite[Eq.
(07.34.21.0088.01)]{Wolfram} and \cite[Eq. (3.381.4)]{Grad.}, we get
\eqref{Eq.14}.

\section{Asymptotic Outage Performance}\label{APA}
At high SNR values, we have
$P_{\sf out}$ $\simeq(G_{c}$SNR)$^{-G_{d}}$, where $G_{c}$ and $G_{d}$ are the system coding gain and diversity order, respectively \cite{Salhab_IEEE_JLT_2016}. Here, the CDF in \eqref{Eq.7} becomes as \cite{Salhab_IEEE_JLT_2016}
\begin{align}\label{Eq.7ew}
F_{\gamma_{\sf D}}(\gamma)&\simeq F_{\gamma_{{\sf
U}_{\mathrm{Sel}},{\sf R}}}(\gamma)+F_{\gamma_{{\sf
R},{\sf D}}}(\gamma).
\end{align}

\subsubsection{First Hop Link}At high average SNR values ($\bar{\gamma}_{u,{\sf r}}\rightarrow\infty$), the CDF of a source channel SNR is approximated by
\begin{align}\label{sese1}
F_{\gamma_{{\sf U}_{u},{\sf R}}}(\gamma)\simeq
\frac{\gamma^{N}}{(C\bar{\gamma}_{u,{\sf r}})^{N} N!}.
\end{align}
Now, upon substituting \eqref{sese1} in \eqref{Eq.8}, we get
\begin{align}\label{dd.1oci1qqqq}
F_{\gamma_{{\sf U}_{\mathrm{Sel}},{\sf
R}}}(\gamma)\simeq
\frac{\gamma^{KN}}{(C\bar{\gamma}_{u,{\sf r}})^{KN} (N!)^{K}}.
\end{align}

\subsubsection{Second Hop Link} From \cite[Eq.
(07.34.06.006.01)]{Wolfram}, as $\bar{\gamma}_{{\sf r},{\sf
d}}$ $\rightarrow$ $\infty$, or equivalently as $z$
$\rightarrow$ $0$, the Meijer G-function can be approximated using
the following series representation
\begin{align}\label{Eq.24}
\mathrm{G}^{m,n}_{p,q}\left[z \bigg|\substack{a_{1},\ldots,a_{p}\\
\\b_{1},\ldots,b_{q}}\right]=&\sum_{k=1}^{m}\frac{\prod^{m}_{\substack{j=1,
j\neq
k}}\Gamma(\delta_{1})\prod^{n}_{j=1}\Gamma(1-\delta_{2})}{\prod^{p}_{j=n+1}\Gamma(\delta_{3})
\prod^{q}_{j=m+1}\Gamma(1-\delta_{4})}\nonumber\\
&\times z^{b_{k}}(1+o(z)),
\end{align}
where $\delta_{1}=b_{j}-b_{k}$, $\delta_{2}=a_{j}+b_{k}$, $\delta_{3}=a_{j}-b_{k}$, $\delta_{4}=b_{j}+b_{k}$, and $p\leq q$ is required. Defining
$\nu=\mathrm{min}\{\alpha,\beta,\zeta^{2}\}$, we get
\begin{align}\label{Eq.25}
F_{\gamma_{{\sf R},{\sf
D}}}(\gamma)\simeq\Upsilon\left(\frac{\gamma}{\bar{\gamma}_{{\sf
r},{\sf d}}}\right)^{\frac{\nu}{r}},
\end{align}
where $\Upsilon$ is constant. Upon substituting
\eqref{dd.1oci1qqqq} and \eqref{Eq.24} in \eqref{Eq.7ew}, the asymptotic outage probability can be written as
\begin{align}\label{Eq.21asa}
&P_{\sf out}^{\infty}=\left(\frac{C}{\gamma_{\sf out}(N!)^{-N^{-1}}}\bar{\gamma}_{u,{\sf r}}\right)^{-KN}+\left(\frac{\Upsilon^{-\frac{r}{\nu}}}{\gamma_{\sf
out}}\bar{\gamma}_{{\sf r},{\sf
d}}\right)^{-\frac{\nu}{r}}.
\end{align}

It is clear from \eqref{Eq.21asa} that the performance of
the considered system is dominated by the worst hop, which depends on the parameters of these two hops. Therefore,
the diversity order $G_{d}$ is equal to
$\mathrm{min}(KN,\frac{\nu}{r})$ and based on
this value, the system performance could be dominated by either:
1) the first hop (i.e. $K$ and $N$) when it is worst, 2) or the second hop (i.e. $\alpha$, $\beta$, $\zeta^{2},\ \mathrm{and}\ r$) when it is worst.

\section{Simulation and Numerical Results}\label{SNRs}
\begin{figure}[htb!]
\centering
\includegraphics[scale=0.37]{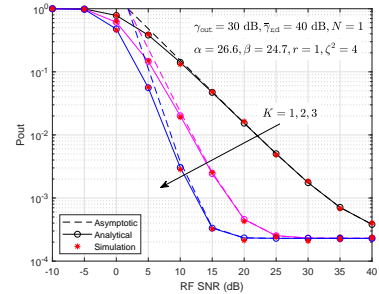}
\caption{$P_{\sf out}$ vs SNR for different values of
$K$.}\label{Pout_SNR_K}
\end{figure}

A good matching between the derived results and simulations is clear in Fig. \ref{Pout_SNR_K}. It is obvious that as $K$ increases, the diversity order $G_{d}$ increases and better the achieved performance. This is expected as when the first hop dominates the performance, $G_{d}=KN$. As $\bar{\gamma}_{\sf r,d}$ is kept constant, a noise floor appears in the results when $\bar{\gamma}_{u,{\sf r}}$ becomes close to or larger than $\bar{\gamma}_{\sf r,d}$. This is because in this region, the second hop becomes dominant and no gain is achieved in the performance when $\bar{\gamma}_{u,{\sf r}}$ keeps increasing. All the curves converge to the same performance in this region as they have no effect here.

\begin{figure}[t]
\centering
\includegraphics[scale=0.37]{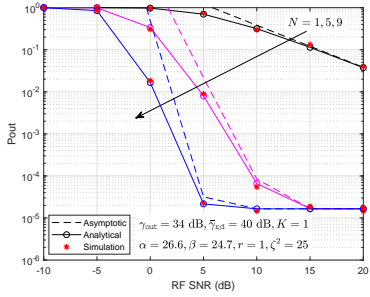}
\caption{$P_{\sf out}$ vs SNR for different values of
$N$.}\label{Pout_SNR_N}
\end{figure}

Fig. \ref{Pout_SNR_N} shows that as $N$ increases, $G_{d}$ increases and better the achieved performance. To study the impact of $N$, $\bar{\gamma}_{\sf r,d}$ is kept constant here. This also results in a noise floor in the results, where when $\bar{\gamma}_{u,{\sf r}}$ becomes close to or larger than $\bar{\gamma}_{\sf r,d}$, no gain is achieved in the performance as the second hop becomes dominant. Accordingly, all the curves converge to the same performance in this region as they have no effect here.

\begin{figure}[t]
\centering
\includegraphics[scale=0.37]{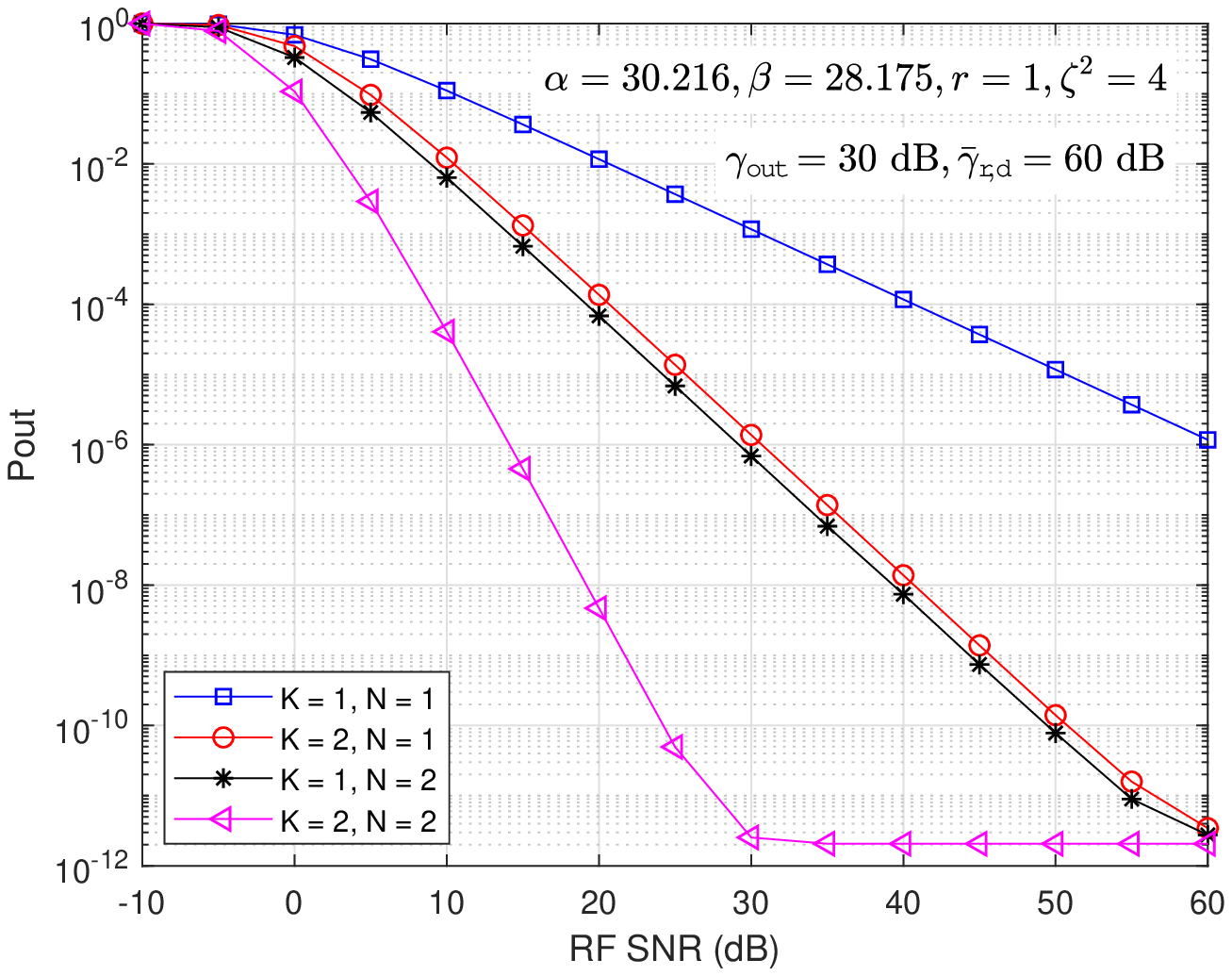}
\caption{$P_{\sf out}$ vs SNR for different values of
$K$ and $N$.}\label{Pout_SNR_K_N}
\end{figure}

\begin{figure}[t]
\centering
\includegraphics[scale=0.37]{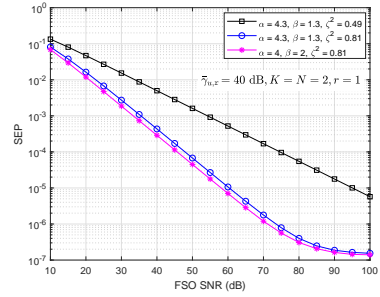}
\caption{ASEP vs SNR for different values of
$\alpha$, $\beta$, and $\zeta^{2}$.}\label{SEP_SNR_alpha_beta_zeta}
\end{figure}

\begin{figure}[t]
\centering
\includegraphics[scale=0.37]{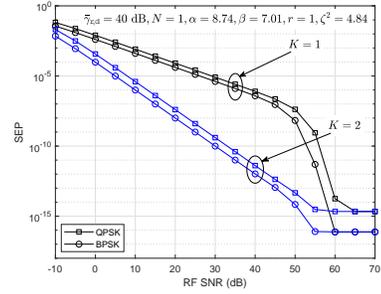}
\caption{ASEP vs SNR for different modulation schemes.}\label{SEP_SNR_K_MSs}
\end{figure}
Fig. \ref{Pout_SNR_K_N} compares the performance under different values of $K$ and $N$. It is obvious that when ($K=N=1$), $G_{d}=1$. For the cases where ($K=2, N=1$) and ($K=1, N=2$), $G_{d}=2$, but the later gives better results in terms of coding gain $G_{c}$. This informs us that one source with two reflecting elements outperforms two sources each of one reflecting element. When ($K=N=2$), $G_{d}=4$ and the best performance is achieved. A noise floor appears as $\bar{\gamma}_{\sf r,d}$ is kept constant here.

The ASEP is studied in Fig. \ref{SEP_SNR_alpha_beta_zeta} for BPSK modulation scheme ($p=q=1$). $\bar{\gamma}_{u,{\sf r}}$ is kept constant and $r=1$. Clearly, when the FSO link's minimum parameter is increased ($\zeta^{2}$), $G_{d}$ increases and the performance is more enhanced as increasing $\zeta^{2}$ reduces the alignment error. On the other hand, when $\zeta^{2}=0.9$ is kept constant and $\alpha$ and $\beta$ are increased, only $G_{c}$ is enhanced. This is because $G_{d}$, when the FSO link is dominant, $G_{d}=\mathrm{min}(\alpha,\beta,\zeta^{2})$. Again, as $\bar{\gamma}_{u,{\sf r}}$ is kept constant here, a noise floor appears in the results when $\bar{\gamma}_{u,{\sf r}}$ becomes close to or larger than $\bar{\gamma}_{\sf r,d}$, as expected.

The ASEP for two modulation schemes is shown in Fig. \ref{SEP_SNR_K_MSs}: BPSK ($p=q=1$) and QPSK ($p=1,q=0.5$). We can see that the QPSK is outperforming the BPSK, as expected. This impact comes in terms $G_{c}$. In addition, the impact of increasing $K$ is clear here through increasing $G_{d}$.


%
\section{Conclusion}\label{C}
Closed-form analytical approximations were derived 
for the outage probability and ASEP of RIS-assisted sources mixed RF/FSO relay network. 
Moreover, the system was studied at the SNR regime whereby $G_{d}$ and $G_{c}$ were
provided. The results illustrated that the system performance is
dominated by the worst hop and that $G_{d}$ is equal to $\mathrm{min}(KN,\alpha,\beta,\zeta^{2})$. In addition, findings showed that for the same $G_{d}$, $N$ is more impactful on the system performance than $K$ through $G_{c}$.







\ifCLASSOPTIONcaptionsoff
\fi


\begin{thebibliography}{1}

\bibitem{Kedar}
D. Kedar and S. Arnon, "Urban optical wireless communications
networks: The main challenges and possible solutions," \emph{IEEE
Commun. Mag.}, vol. 42, no. 5, pp. 2--7, 2003.

\bibitem{Laneman}
J. N. Laneman, D. N. C. Tse and G. W. Wornell, "Cooperative
diversity in wireless networks: Efficient protocols and outage
behavior," \emph{IEEE Trans. Info. Theory}, vol. 50, no. 12, pp.
3062--3080, 2004.

\bibitem{Lee}
E. Lee, J. Park, D. Han, and G. Yoon, "Performance analysis of the
asymmetric dual-hop relay transmission with mixed RF/FSO links,"
\emph{IEEE Photon. Technol. Lett.}, vol. 23, no. 21, pp.
1642--1644, 2011.

\bibitem{Ansari1}
I. S. Ansari, F. Yilmaz, and M.-S. Alouini, "Impact of pointing
errors on the performance of mixed RF/FSO dual-hop transmission
systems," \emph{IEEE Wireless Commun. Lett.}, vol. 2, no. 3, pp.
351--354, 2013.

\bibitem{Renzo1}
M. D. Renzo \textit{et al.}, "Reconfigurable intelligent surfaces vs.
relaying: Differences, similarities,
and performance comparison," \textit{IEEE Access}, vol. 1, pp. 798--807, 2020.

\bibitem{Basar1}
E. Basar, M. D. Renzo, J. D. Rosny, M. Debbah, M.-S. Alouini, and
R. Zhang, "Wireless communications through reconfigurable intelligent
surfaces," \textit{IEEE Access}, vol. 7, pp. 116753--116773, 2019.

\bibitem{Wu}
Q. Wu, and R. Zhang, "Intelligent reflecting surface enhanced wireless
network via joint active and passive beamforming," \textit{IEEE Trans. Wireless
Commun.}, vol. 18, no. 11, pp. 5394--5409, 2019.

\bibitem{Yang3}
L. Yang, W. Guo, and I. S. Ansari, "Mixed dual-hop FSO-RF communication
systems through reconfigurable intelligent surface," \textit{IEEE Commun.
Lett.}, vol. 24, no. 7, pp. 1558--1562, 2020.

\bibitem{Yang6}
L. Yang, F. Meng, Q. Wu, D. B. D. Costa, and M.-S. Alouini, "Accurate closed-form approximations to channel
distributions of RIS-aided wireless systems," \textit{IEEE Wireless Commun. Lett.}, vol. 9, no. 11, pp. 1985--1989, 2020.

\bibitem{Tang}
W. Tang, J. Y. Dai, M. Chen, X. Li, Q. Cheng, S. Jin, K.-K. Wong, and
T. J. Cui, "Programmable metasurface-based RF chain-free 8PSK wireless
transmitter," \textit{Electron. Lett.}, vol. 55, no. 7, pp. 417--420, 2019.

\bibitem{Salhab_IEEE_JLT_2016}
A. M. Salhab, F. Al-Qahtani, R. M. Radaydeh, S. A. Zummo, and H.
Alnuweiri, "Power allocation and performance of multiuser mixed
RF/FSO relay networks with opportunistic scheduling and outdated
channel information," \emph{IEEE/OSA J. Lightw.
Technol.}, vol 34, no. 13, pp. 3259--3272, 2016.

\bibitem{Zhang}
W. Zhang, S. Hranilovic, and C. Shi, "Soft-switching hybrid FSO/RF
links using short-length raptor codes: Design and implementation,"
\emph{IEEE J. Sel. Areas Commun.}, vol. 27, no. 9, pp. 1698--1708,
2009.

\bibitem{Grad.}
I. S. Gradshteyn and I. M. Ryzhik, \emph{Tables of Integrals,
Series and Products}, 6th ed., San Diago: Acadamic Press, 2000.

\bibitem{Ansari3}
I. S. Ansari, F. Yilmaz, and M.-S. Alouini, "Performance analysis
of free-space optical links over M\'{a}laga (M) turbulence
channels with pointing errors," \emph{IEEE Trans. Wireless
Commun.}, vol. 15, no. 1, pp. 91--102, 2016.

\bibitem{AnasP}
A. M. Salhab, "A new scenario of triple-hop mixed
RF/FSO/RF relay network with generalized
order user scheduling and power allocation," \textit{J. on Wireless Commun. and Netw.}, (2016) 2016:260, 2016.

\bibitem{McKay}
M. R. McKay, A. L. Grant, and I. B. Collings, "Performance
analysis of MIMO-MRC in double-correlated Rayleigh environments,"
\emph{IEEE Trans. Commun.}, vol. 55, pp. 497--507, 2007.

\bibitem{Wolfram}
Wolfram, "The Wolfram functions site," Available:
http://functions.wolfram.com, 2013.


\end{thebibliography}
\end{document}